\documentclass[]{article}

\AtBeginDocument{%
  }

\usepackage{url}
\usepackage{todonotes}
\usepackage{hyperref}

\usepackage{datetime}
\ddmmyyyydate

\usepackage{cleveref}
\usepackage{xspace}

\newcommand*{\detection}{{\emph{Malice Detection}}\xspace}
\newcommand*{\quarantine}{{\emph{Quarantine-Release Criterion}}\xspace}
\newcommand*{\execution}{{\emph{Transaction Execution}}\xspace}

\newcommand*{\name}{\textsc{Zircuit}\xspace}

\usepackage{vhistory}

\begin{document}


\title{Sequencer Level Security}

\author{Martin Derka\\
\texttt{martin@zircuit.com}
\and Jan Gorzny\\
\texttt{jan@zircuit.com}
\and Diego Siqueira\\
\texttt{diego@zircuit.com}
\and Donato Pellegrino\\
\texttt{donato@zircuit.com}
\and Marius Guggenmos\\
\texttt{marius@zircuit.com}
\and Zhiyang Chen\\
\texttt{jeff@zircuit.com}
}

\date{{\bf 1 Dec 2023}\\ {\small Updated May 2024}}



\maketitle

\begin{abstract}
    Current blockchains do not provide any security guarantees to the smart contracts and their users as far as the content of the transactions is concerned. In the spirit of decentralization and censorship resistance, they follow the paradigm of including valid transactions in blocks without any further scrutiny. 
    Rollups are a special kind of blockchains whose primary purpose is to scale the transaction throughput. 
    Many of the existing rollups operate through a centrally operated sequencing protocol. 
    In this paper, we introduce the Sequencer Level Security (SLS) protocol, an enhancement to  sequencing protocols of rollups.  
    This pioneering contribution explores the concept of the sequencer's capability to identify and temporarily quarantine malicious transactions instead of including them in blocks immediately. 
    We describe the mechanics of the protocol for both the transactions submitted to the rollup mempool, as well as transactions originating from Layer one. 
    We comment on topics such as trust and decentralization, and consider the security impact on the protocol itself. 
    We implement a prototype of the SLS protocol, {\name}, which is built on top of Geth and the OP stack. 
    The SLS protocol described can be easily generalized to other rollup designs, and can be used for purposes other than security.
\end{abstract} 

\section{Introduction}\label{sec:intro}

Blockchain technology has revolutionized the way we think about financial transactions and data storage, by offering a decentralized, resilient, and programmable ledger that operates on a global scale.
The introduction of \emph{smart contracts}~\cite{szabo1997idea} on platforms such as Ethereum~\cite{ethereum} has further expanded the capabilities of blockchains, allowing users to execute deterministic programs stored on the blockchain. 
These smart contracts are invoked through transactions, and every node in the network executes the code to update the state. However, despite these innovations, security vulnerabilities have been a persistent threat, leading to significant financial losses within the blockchain ecosystem.

Notably, even the most decentralized blockchains have found themselves compelled to take centralized measures in response to major security breaches. In 2010, a vulnerability in the code of Bitcoin~\cite{bitcoin} led to the creation of a block that contained a transaction creating 184 billion Bitcoin. The community quickly reacted by releasing a patch and introduced a soft fork to nullify the impact of the exploit~\cite{bitcoinincident}. In 2016, 
the DAO protocol on Ethereum was hacked for \$60 million, leading to a contentious hard fork to reverse the exploit~\cite{ethereumdao}. More recently, in 2022, the Binance Smart Chain took the drastic step of pausing and rolling back blocks to undo a cross chain hack which stole \$2 million BNB tokens~\cite{binancehack}. 
These incidents underscore the challenges faced by decentralized blockchains when confronted with significant exploits and the lack of effective mechanisms to preemptively detect, isolate, and block them.


In the realm of Ethereum, rollups, a.k.a~\emph{commit-chains} \cite{cryptoeprint:2018/642} or \emph{validating bridges} \cite{cryptoeprint:2021/1589}, have become a crucial Layer 2 (L2) solution for enhancing transaction throughput on the Layer 1 (L1) Ethereum network (see e.g.,~\cite{GudgeonMRMG20} for more on Layer 2 networks). 
Rollups interpret transactions and maintain a separate blockchain state, but do not typically need to form consensus for this purpose. 
The order of transactions and their assignment to blocks, and consequently the state of the rollup, are decided by a so-called sequencer. 
As the sequencer dictates the order of transactions and blocks, other L2 participants do not need to reach consensus on L2 transactions and blocks.
This results in the capability for the network to produce blocks at a higher rate than Ethereum. Almost all rollups currently have a single centralized sequencer.

\begin{table}[]
    \caption{Layer 2 (L2) Chains and their Sequencer Status. 
    \textbf{TVL}: indicates the Total Value Locked expressed in billions of dollars. 
    \textbf{Single Seq}: denotes whether a single centralized sequencer is used. 
    \textbf{Tx w/o Seq.?}: stands for the ability to submit transactions without the sequencer, where "Yes" is followed by the delay time required for such transactions.
    }
    \label{tab:l2study}
    \centering
    \begin{tabular}{|l|l|l|l|}
    \hline
    L2 Chain     & TVL (\$B) & Single Seq. & Tx w/o Seq.?                            \\ \hline
    Arbitrum One & 18.65    & Yes          & Yes (1d)~\cite{l2beatArbitrum}          \\ \hline
    Optimism     & 7.62     & Yes          & Yes (12h)~\cite{l2beatOptimism}         \\ \hline
    Base         & 4.09     & Yes          & Yes (12h)~\cite{l2beatBase}             \\ \hline
    Blast        & 2.75     & Yes          & Yes (12h)~\cite{l2beatBlast}            \\ \hline
    Starknet     & 1.37     & Yes          & No~\cite{l2beatStarknet}                \\ \hline
    zkSync Era   & 0.746    & Yes          & No~\cite{l2beatZkSyncEra}               \\ \hline
    \end{tabular}
    \end{table}

We conducted a study of the top six rollups ranked by Total Value Locked (TVL), as reported by L2Beat~\cite{l2beat}. 
The TVL was determined by aggregating the value (in USD) of canonically bridged, externally bridged, and natively minted assets~\cite{l2beattvl}.
Table~\ref{tab:l2study} presents the results of our study. 
All six rollups have a single centralized sequencer controlled by the L2 team. Two of them (StarkNet and zkSync Era) do not allow transactions to be submitted without the sequencer. The other four (Arbitrum One, Optimism, Base, and Blast) allow transactions to be submitted without the sequencer, via sending messages on L1, but with a delay of at least 12 hours and higher gas fees. The centralized nature of sequencers in rollups has been a subject of debate within the community. Despite this, the community has largely accepted the centralization of sequencers in rollups, as it has enabled significant improvements in transaction throughput~\cite{centralizedrollup}.

The Ethereum protocol, in the spirit of absolute censorship resistance, does not examine the content of transactions prior to their inclusion in blocks. Essentially, any transaction that pays the necessary gas fee and bears a valid signature gets added to a block. However, community-driven initiatives such as Flashbots~\cite{flashbots} have managed to assemble miners to rearrange the order of transactions within a block, motivated by both profit and security considerations (see Section~\ref{sec:related}). It is important to note that while services like Flashbots can filter out or re-order certain transactions, they cannot catch all malicious activity. Attackers still have the opportunity to commit harmful transactions through miners not participating in Flashbots.

Rollups, much like Ethereum, currently lack mechanisms for scrutinizing transactions before they are included in blocks.
Given that rollups typically operate under a centralized sequencer, it is also impossible for the community to develop protocols similar as Flashbots on rollups in the future. This situation presents a unique opportunity: what if we could equip the sequencer with an extra layer of security? 
This layer's primary motivation is not profit, but solely enhancing security by preventing malicious transactions from being executed.

Recognizing the critical need for improved security measures, we propose the Sequencer Level Security (SLS) protocol. This novel approach leverages the centralized nature of sequencers to scrutinize transactions for potential malicious intent before they are finalized on Layer 2. By enabling the early detection and quarantine of suspect transactions, the SLS protocol aims to enhance the security of smart contracts and the underlying blockchain without necessitating the contentious measures of hard forks or block reversion.

\noindent \textbf{Contributions:}  This paper presents the following contributions:

\begin{itemize}
    \item \textbf{Sequencer Level Security (SLS):} This paper introduces a novel approach to enhancing transaction security on rollups with centralized sequencers. SLS enables early detection and quarantine of potentially malicious transactions, augmenting the blockchain's security framework.
    \item \textbf{SLS Customization and Trust Assurance:}  SLS enables users to set customized invariants for malice detection, allowing more granular control over security protocols. It also permits users to stake assets to  the sensitivity of malice detection, facilitating the processing of transactions with acceptable risk levels.
    \item \textbf{Implementation of SLS:}  We detail the implementation of the SLS protocol (called {\name}) atop the so-called OP stack, an open source implementation of the Optimism rollup \cite{opstack}. 
    This also includes comprehensive insights into the design choices, technical nuances, and the integration process, providing a blueprint for deploying SLS in real-world scenarios. 
    \item \textbf{Critical Insights and Discussion on SLS:} Our research offers critical insights into the potential impacts, challenges, and future directions of the SLS protocol. We discuss the implications of our findings for the development and application of security protocols in decentralized networks, setting the stage for further innovations in the field.
\end{itemize}

The paper is organized as follows: Following this introduction, we present the preliminaries of Ethereum and the Optimism\footnote{``Bedrock'' release} architecture in Section~\ref{sec:prelim}. Section~\ref{sec:sls} introduces our main contribution, the SLS protocol, and Section~\ref{sec:implementation} discusses the implementation details and concerns. 
Section~\ref{sec:related} provides an overview of related work, and Section~\ref{sec:conclusion} concludes the paper. 
\section{Preliminaries}\label{sec:prelim}

A blockchain serves as a distributed ledger across a computer network, where nodes use consensus algorithms to appoint a leader for appending data. 
Each node maintains a full blockchain copy, validating transactions and blocks to ensure integrity and continuity. 
A notable example is the Ethereum blockchain, which we will use throughout this paper without loss of generality.
This section provides the necessary background on Layer 1 and Layer 2 networks, the Ethereum blockchain, and the Optimism rollup stack for the rest of the paper.




\subsection{Layer 1 (L1) and Layer 2 (L2) Networks}\label{subsec:l1l2}

Blockchains like Ethereum offer a decentralized framework allowing mutually mistrusting entities to cooperate without a trusted third party. 
Despite their transformative potential, these blockchains are constrained by limited throughput. 
The academic and practical discourse on blockchain scalability has produced several approaches, including sharding~\cite{zamani2018rapidchain} and side-chains~\cite{back2014enabling}.

Layer 2 (L2) protocols represent an orthogonal scaling solution. Unlike the aforementioned approaches, Layer 2 protocols enhance scalability without altering the trust assumptions or consensus mechanisms of the blockchain to be scaled, referred to as \textbf{Layer 1 (L1)}. 
L2 enables users to conduct transactions off-chain through private and authenticated communication, avoiding the need to broadcast every transaction on the parent blockchain~\cite{GudgeonMRMG20}. 
These off-chain transactions effectively form another blockchain, also termed \textbf{Layer 2 (L2) blockchain}. 
For instance, Optimism serves as a L2 blockchain that scales the ecosystem of Ethereum, its corresponding L1 blockchain.

\subsection{Ethereum Protocol Overview}\label{subsec:ethereum}

Ethereum's architecture comprises a network of nodes, each upholding the blockchain's state. 
They form consensus on who the next leader allowed to propose a block through the Proof-of-Stake protocol (see e.g., \cite{8746079}).
The leader forms a block filled with transactions that is broadcasted to the rest of the network nodes. The nodes confirm the validity of transactions, execute the transactions, and update their blockchain state. This way, the blockchain forms consensus on the state itself. 

The transactions contained in the blocks are submitted to the nodes by the blockchain users. The users are represented by Externally Owned Accounts (EOA).
Each EOA is represented by an address that is derived from the private key that the account uses to sign its transactions. The nodes maintain the transactions submitted to them yet not included in the blocks in so-called \textbf{mempool}. 
The transactions submitted to the mempool are propagated among the nodes over a P2P~\cite{schollmeier2001definition} protocol so that they can be included in the blocks by any node selected as the next block proposer.

Transactions submitted by users must meet specific criteria and be properly signed. In order to ensure proper sequential processing, each account needs to tag its transactions with so-called \textbf{nonce}. 
The nonce is a number that determines the sequential order of the transaction among the other transactions signed and submitted by the same account. Nonces for each account start with 0. 
The Ethereum nodes will accept and process account's transactions only sequentially, and the protocol does not allow any gaps. 
In most implementations, if a node receives two transactions with the same nonce signed by the same account in the mempool, it retains the transaction with higher gas price and drops the transaction with the lower gas price (the concept of gas is described later in this section). This mechanic can be used for replacing transactions that were broadcasted to the network and were not included in blocks yet. Once a transaction is included in a block, the inclusion is permanent, which guarantees immutability of the blockchains state.

The Ethereum network supports \textbf{smart contracts}, which are small deterministic programs stored on the blockchain that users can invoke through transactions. Each smart contract is represented by its own unique account. Unlike externally owned accounts (EOAs) that are controlled by users via private keys, smart contracts operate differently; they do not have associated private keys in the conventional sense. This structural distinction ensures that, assuming the underlying cryptography remains secure, it is practically impossible to externally control a smart contract without explicit permission encoded within its functions. Moreover, while transactions submitted to the network must be signed by EOAs, indicating that smart contracts cannot initiate transactions on their own, interactions between smart contracts within a transaction are both possible and common.

\textbf{Ether} (Eth) is the virtual currency that the Ethereum network uses to charge costs related to the computing power required to process transactions. 
As part of the blockchains state, Ethereum maintains the balance for each account. 
The cost of processing transactions is expressed so-called \textbf{gas} units—every operation carries a cost that contributes to the final price. When accounts submit transactions, they choose how much Ether they are willing to pay per unit of gas. 
This price is divided into two parts—a \textbf{base fee} and \textbf{priority fee}. The base fee stipulates the minimum that the account needs to pay for a unit of gas. 
The priority fee represents how much more Ether the account is willing to pay for including the transaction in a block with priority. 
When the transaction becomes processed and included in the block, the balance of the account that signed decreases by the respective amount Ether, the portion corresponding to the base fee is removed from circulation (the usual jargon for this action says that the Ether is ``burnt'' or ``burned''), and the portion corresponding to the priority fee is given to the proposer of the block. 

Transactions on Ethereum are uniquely identified by hashes. 
A transaction hash is calculated based on the nonce, sender, recipient, transaction data, and value.
Ethereum uses a modified Merkle Patricia Trie (MPT)~\cite{mpt} data structure to efficiently store and retrieve the state. 
The entire state of Ethereum (all accounts and their respective states) is stored in this trie. 
The state root is a 256-bit hash which uniquely identifies the root of the MPT containing all the state information.

The logic of the Ethereum network described in this section is driven by the so-called Ethereum Virtual Machine (EVM).
Other blockchain networks that use the same logic as EVM are called EVM-compatible. 
Go-Ethereum, or simply \textbf{Geth}~\cite{geth}, is a popular implementation of the Ethereum network client.

\subsection{Optimism Protocol Overview}\label{sec:optimism}

Optimism publishes transaction data and blockchain output roots onto Ethereum.
By observing this data, one can recreate the Optimism blockchain state entirely (assuming that no invalid output root updates were provided and successfully challenged; see e.g. \cite{optimismchallenge}).

The Optimism network consists of Optimism nodes, a special privileged node called the sequencer, and  L1 smart contracts.

A node is composed of the custom Optimism network derivation software, the ~\texttt{op-node} service, and a custom version of Geth, the so-called \texttt{op-geth} service, as well as a connection to the underlying L1. 
The \texttt{op-node} provides info about the current tip of the L2 blockchain and performs the derivation of the entire blockchain by reading the L2 transactions published on the L1.
The \texttt{op-node} replays the Optimism transactions when operating in replica (to recreate the state locally) or verifier modes (to potentially challenge a state update), or for the first time if the node is the sequencer. 
The transactions are executed via \texttt{op-geth}, which is a modified version of Geth to support some additional functionality.
For example, \textbf{deposit transactions} (defined below) are those that originate on L1 but must be included on L2; this functionality does not exist for the closed-ecosystem blockchain that Geth normally implements.

The \textbf{sequencer} is a privileged node that collects users' transactions similar to Ethereum and also order them based on pre-defined rules.
Transactions are kept in a mempool and propagated via P2P to the network of nodes. 
The sequencer's set of components additionally includes a batcher service and a proposer service. 
The batcher service publishes the L2 transaction data in form of batches on the L1 as \texttt{calldata}.\footnote{After EIP-4844~\cite{eip4844-protodanksharding} was introduced in Ethereum Cancun Upgrade on March 13, 2024, the batcher service can also publish the L2 transaction data as blobs on the L1.}
This data is posted as \texttt{calldata} on L1 because it is not explicitly used as program inputs, but is simply recorded there for others to observe. 
This is the data that is used by the other nodes to derive the chain of blocks and determine the state of the L2.
As Optimism relies totally on a single sequencer for transaction inclusion and ordering, the batches posted onto Ethereum by the batcher which persist for a sufficient number of L1 confirmations, are therefore those that are (eventually) executed on the Optimism network.
The proposer service publishes the new state roots generated by the execution of the blocks containing the batched transactions. 
An output root is the hash of data including the block hash, the L2 state root and the storage hash of the \texttt{L2ToL1MessagePasser} smart contract on the L2. 

The Optimism L1 smart contracts are used to perform both deposit transactions (simply ``desposits'') and withdrawal transactions (simply ``withdrawals'') from the L2 network, as well as to verify the state roots.

A \textbf{deposit transaction} is an L2 transaction generated by an L1 transaction. 
Deposits can be used to credit Eth on an L2 address but also to include an arbitrary transaction on the L2 (in this sense, the name ``deposit'' is not entirely accurate). 
In particular, they can be used to perform any smart contract interaction such as transferring tokens. 
When a user performs a deposit calling the relevant smart contract (\texttt{OptimismPortal}), the smart contract emits an \texttt{event}.
This \texttt{event} contains all the transaction parameters such as sender, recipient, value and data. Deposits must be included in the derivation of the chain by the protocol and are for this reason censorship resistant.

The verification of the state roots is done by the Fault Dispute Game that is currently under development \cite{disputegame}. 
State roots are published on the L1 by the sequencer (namely the proposer) and are optimistically considered valid if they have not been challenged for the duration of a challenge period; 7 days in Optimism. 
The challenge game starts when a state root is challenged by a node that deems it invalid. 
The game consists in performing a binary search over intermediate states until two consecutive states are found. 
The second state is the challenged one. 
At this point the verification of the validity of the second state is performed by a smart contract that can execute MIPS~\cite{kane1992mips} bytecode. 
The \texttt{op-node} code in Go~\cite{meyerson2014go} can be compiled in MIPS and executed onchain. 
The smart contract executes the instruction that generates the second state and decides if the state root corresponding to the current game is invalid or not. 
If the state root is not invalid, it will be considered final if it remains unchallenged for one week. 
The game is accompanied by a set of incentives based on staking: if a state is proven invalid, the proposer's stake is forfeited.

Withdrawals depend on state roots, and hence they can be performed only after a state root is considered valid. 
Withdrawals are initiated on the L2 by transacting on the appropriate L1 smart contract (\texttt{L2ToL1MessagePasser}). 
First, the contract hashes the transactions parameters and saves the hash in an array in its storage. 
This is called a \emph{withdrawal initiating transaction}. 
On the \texttt{OptimismPortal} on the L1, once the output root is proposed, a user (or relayer) can perform a withdrawal proving transaction: this transaction proves that a transaction is included in a state root by providing a merkle proof. 
After the state root is considered valid, a user can perform a withdrawal finalising transaction that effectively executes the withdrawal. 
A withdrawal lets users withdraw Eth from the \texttt{OptimismPortal} and execute any L2-to-L1 transaction.

\section{Sequencer Level Security (SLS)}\label{sec:sls}

\begin{figure*}[!htbp]
    \centering
    \includegraphics[width=0.8\linewidth]{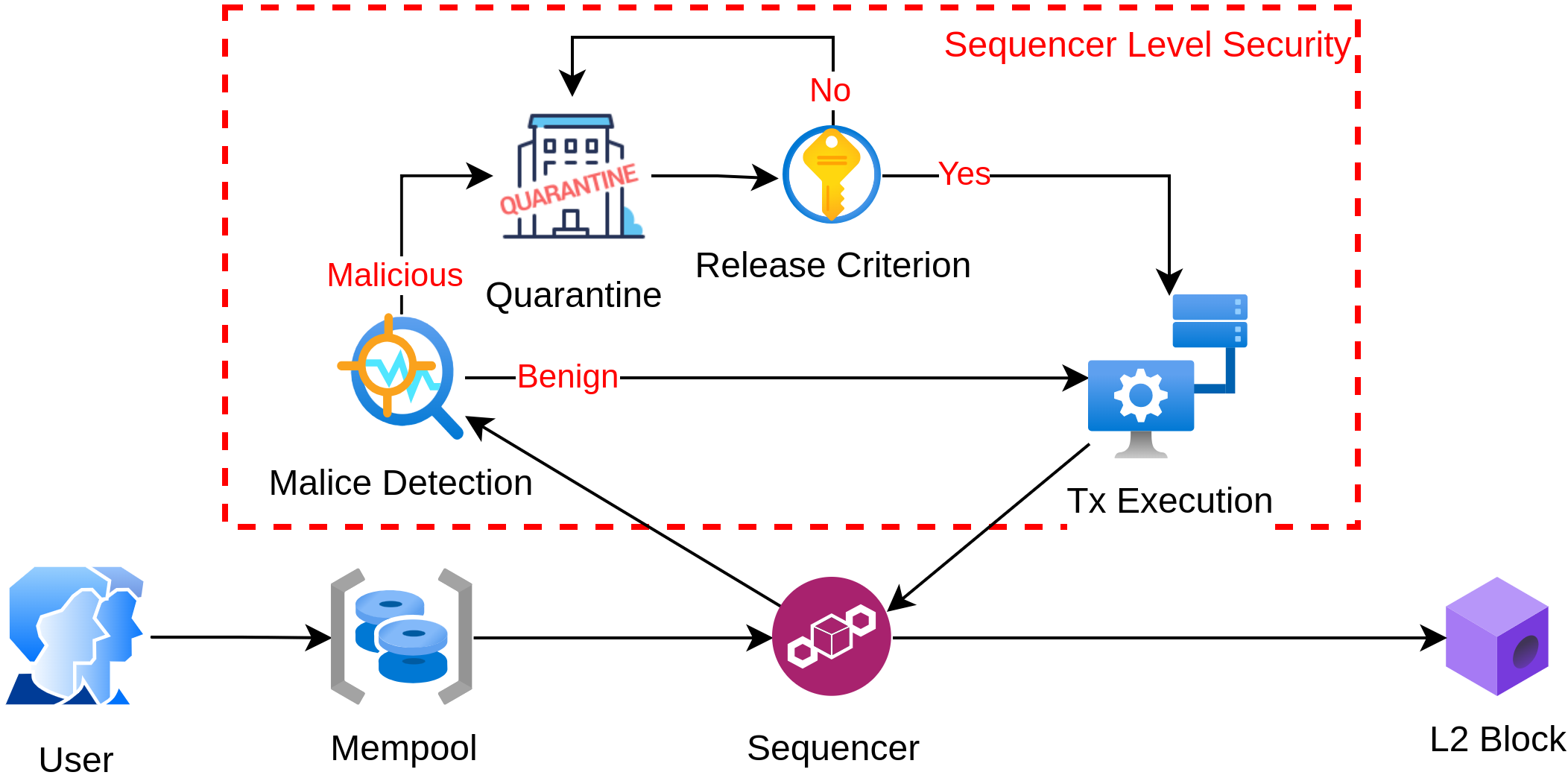}
    \caption{An Overview of the Sequencer Level Security (SLS) Protocol.}
    \label{fig:SLS}
\end{figure*}

In this section we introduce Sequencer Level Security, a sequencing protocol that provides additional enhanced security guarantees compared to Ethereum and the current state-of-the-art rollups. 

Figure~\ref{fig:SLS} presents an overview of the protocol. It contains three main components: (1) \detection, (2) \quarantine, and (3) \execution. 

Upon arrival at the SLS sequencer, transactions from the mempool are initially routed to \detection module. It identifies whether a transaction is benign or potentially malicious (detailed in Section \ref{sec:malice-detection}).
Benign transactions are promptly queued for block inclusion, adhering to standard sequencing protocols. Conversely, transactions flagged as malicious are diverted to \quarantine module, which acts as an intermediary holding area. Here, they undergo a rigorous verification process against specific release criteria (detailed in Section \ref{sec:quarantine}). Transactions that meet these criteria are then forwarded to \execution module.
\execution module executes the transactions against the blockchain state at the forthcoming L2 block. Successfully executed transactions are cycled back to the SLS sequencer for inclusion in the forthcoming L2 block (detailed in Section \ref{sec:transaction-execution}).

\subsection{Assumptions and Design Goals}
Throughout this section, we make the assumption that the sequencer is centralized but trusted. Note this assumption is currently valid for almost all popular rollups as shown in Table~\ref{tab:l2study}. 

We further assume that malicious actors are incapable of compromising the sequencer. Specifically, it is presumed that these actors do not have control over the sequencer and cannot exploit it to their advantage. Instead, their interaction with the blockchain is limited to the transactions they submit. Any attempt to utilize blockchain or smart contract vulnerabilities to manipulate the system or illicitly access resources is deemed malicious activity.

We aim to design a protocol that 
\begin{itemize}
    \item minimally impacts sequencing throughput or latency;
    \item is transparent and allows for deterministic derivation of the chain's state; and
    \item maintains, without worsening, the trust model of the existing sequencer for benign actors. 
\end{itemize}

\subsection{Malice Detection}\label{sec:malice-detection}
When the SLS sequencer receives transactions in the mempool, prior to including them in the block, it decides whether the transactions are malicious or not. 
Benign transactions can be considered for immediate inclusion into blocks according to the regular sequencing rules. 
Malicious transactions are placed in the quarantine and cannot be considered for block inclusion until they are released (see Section \ref{sec:quarantine}).

The specific algorithm used by the sequencer to identify malicious transactions is beyond the scope of this paper. 
Possible approaches include program analysis and machine learning(See Section~\ref{subsec:malicious}). 
In particular, tools like Trace2Inv \cite{chen2024demystifying} allow contract deployers to specify invariants that should hold for all transactions, which provides contract deployer customizable choices for the detection of malicious transactions.
Our focus is on the broader mechanics of transaction processing and block formation. Therefore, we assume that the sequencer effectively identifies and segregates malicious transactions, without delving into the algorithmic nuances of this process. 
The minimal usable information that the analysis needs to return is a boolean value indicating whether the transaction should be quarantined or not. 
However, additional output information can be useful for specific release criteria (see the Economic criterion in Section \ref{sec:quarantine}). 

Transaction malice detection typically requires simulation results of a transaction. Such simulation can be performed in two ways:

\noindent \textbf{Isolated Simulation at the Chain's Tip:} Transactions can be simulated independently at a blockchain states of the next block, known as the chain's tip. Here, all transactions are executed using the same state of the blockchain, which ensures that the simulation outcomes pertain solely to individual transactions without the influence of others. This method allows for parallel processing, significantly speeding up the analysis. However, because these simulations lack the context of the block into which transactions would ultimately be integrated, there is a risk of overlooking some malicious activities.

\noindent \textbf{Contextual Simulation within the Block:} Alternatively, transactions can be simulated in the context of their specific block. This process involves sequentially simulating each transaction, with the state of the blockchain being updated by each preceding transaction. While this approach provides a comprehensive view of how transactions influence one another and adapts to each new state, its sequential nature prevents parallel processing, making it slower but more thorough in identifying potential anomalies.

Simulating every transaction at the tip of the chain can result in false positives and false negatives.
For instance, imagine a Decentralized Finanaice (DeFi) protocol $P$ (see e.g., \cite{DBLP:conf/aft/Werner0GKHK22}) that implements a pausing feature, a transaction $A$ that unpauses $P$, and a transaction $B$ that can exploit $P$ provided that $P$ is not paused, 
the sequence of transactions $(A,B)$ would result in an exploit. 
Transaction $B$ alone would not result in an exploit if it was simulated on the tip of the chain when $P$ is paused, and might be considered benign. 
If the sequencer subsequently placed the transaction in a block after $P$ is unpaused, the exploit would occur despite the judgment that $B$ is benign. 
Overall, assessing transactions in isolation at the tip of the chain can yield both false positives and false negatives.

\subsubsection{Hybrid Parallel-Sequential Malice Detection}
We now present an algorithm that combines the two approaches to optimize the speed of detection but maintain its accuracy. 

\noindent \textbf{Step 0 (Choosing Transactions by the Sequencer):} The sequencer selects a list of transactions ($T_1, T_2, ..., T_n$) for potential inclusion in the upcoming block. This step is same as other standard sequencing protocols.

\noindent \textbf{Step 1 (Parallel Simulation on the Tip of the Chain):} Each transaction $T_i$ is independently simulated using the current state at the tip of the blockchain. This step allows for parallel processing of transactions. The outcomes of these simulations provide essential data for future dependency analysis and malice detection: (1) Simulation results of each transaction $T_i$, (2) The blockchain states read ($R_i$) and written ($W_i$) by each transaction $T_i$.

\noindent \textbf{Step 2 (Transaction Dependency Analysis):} For any $i < j$, a transaction $T_j$ is \textbf{dependent} on $T_i$ if $W_i \cap R_j \neq \emptyset$. Informally, a transaction $T_j$ is dependent on $T_i$ if executing $T_i$ may change the outcome of executing $T_j$. This step identifies dependencies among all transactions.

\noindent \textbf{Step 3 (Parallel Detection for Independent Transactions and Sequential Detection for Dependent Transactions):} For any transaction $T_j$ that is not dependent (a.k.a. independent) on any other prior transaction $T_i (i < j)$, the sequencer can perform parallel detection on their simulation results. Other dependent transactions are queued for sequential simulation and detection within the block context. 

\noindent \textbf{Step 4 (Transaction Inclusion):} The sequencer finalizes the block by including all transactions identified as benign in Step 3. Dependent transactions that could not be fully analyzed due to time constraints or complexity are deferred to the next cycle. The same detection process will be applied in the next round when these transactions are considered again for inclusion.

This Hybrid Parallel-Sequential Malice Detection algorithm ensures a robust approach by combining both isolated and contextual analyses of transactions, optimizing for both speed and accuracy in the detection process. 

When a transaction $T_i$, is identified as malicious, it is quarantined by the sequencer. These transactions may only be reconsidered for inclusion and re-simulated only if certain parties subjectively trigger the release criteria, as described in Section \ref{sec:quarantine}. The sequencer conducts only basic checks (a.k.a. retirement criteria), such as verifying the nonce and time criteria, to determine if these transactions can be removed from the mempool. This strategy reduces the computational demand on the sequencer and enhances its defense against Denial-of-Service attacks (see Section~\ref{subsec:risks}).

\subsection{Quarantine}\label{sec:quarantine}
While in quarantine, the transaction is present in the mempool, but it does not change status. Explicitly, it does not get executed and cannot be included in the blocks. The sequencer maintains the information about when the transaction has been placed in the quarantine. The transaction will either be dropped from the mempool once it meets one of the retirement criteria or be released from the quarantine if it meets one of the release criteria. Upon release, it can be selected for block inclusion based on the standard algorithm implemented by the sequencer. 

The exact retirement criteria and release criteria can be defined by the sequencer. The following list of retirement and release criteria appears viable from the security standpoint.

\noindent \textbf{Retirement Criteria:}
\begin{itemize}
\item Nonce criterion. If the transaction can no longer be included in a block because the nonce is no longer valid (i.e., the account has submitted another transaction with the same nonce that has been included in a block), the transaction can be released from the quarantine as it will be subsequently dropped from the mempool.
\item Time criterion. The time criterion represents the reaction time that the sequencer offers to the users to react to a malicious transaction. 
If the transaction has been quarantined for longer than required, it can be released and considered for block inclusion. The exact amount of time required for the transaction to stay in the quarantine is a configuration parameter that can be decided by the blockchain.  
\end{itemize}

\noindent \textbf{Release Criteria:}

\begin{itemize}
\item Failure criterion. If a transaction fails due to changes in the chain's state, it can be safely included in the block since it will result in a revert. Reverted transactions do not alter the blockchain state. Consequently, these transactions are inherently benign.
\item Administrative criterion. It is expected that the detection of malice will occasionally produce false positives. Under such circumstances, the sequencer operational team, comprising security experts, can administratively override decisions to release transactions. This can be executed either through manual review by security experts or an automated agent. Additionally, the malice detection approaches could help identify contracts at risk of financial loss. If identified, these potential victim contract administrators have the option to assess the risks and approve the transaction. Once all necessary approvals are obtained, the transaction is released from quarantine, ensuring a balance between security and operational fluidity.
\item Economic criterion. For this criterion, the sequencer can offer the accounts an option to provide collateral that can be slashed in case the transaction causes harm on-chain. The economic criterion is met if the available collateral staked by the account submitting the transaction exceeds the maximum possible damage that the sequencer anticipates the transaction to cause.
\end{itemize}

While the transaction is in quarantine the sequencer periodically checks whether one of the retirement criteria has been satisfied with every new block appended to the chain. The sequencer does not need to vet the transactions more often as without a block, the state of the chain does not change. The sequencer does not check release criteria, instead, it waits for a privileged party to trigger the release criteria, such as approvals from security experts or collaterals submitted by transaction originators. 

The transaction hash serves as an identifier for transactions in the quarantine, however, the SLS protocol also has to use the account address, and the transaction data (the function selector and call data), and value, to establish whether a newly incoming transactions is a duplicate of a transaction that has been already released from the quarantine (see Section \ref{sec:transaction-execution} for more details). 

While quarantined, the transaction is subjected to the regular retirement criteria that the sequencing protocol prescribes for the mempool. If the transaction retires from the mempool prior to its release from quarantine, it should be removed from quarantine as if it never entered. 
Explicitly stated, in the case of retirement from the mempool, if a transaction with the same hash enters the mempool again (potentially due to re-propagation by the network) and is still considered malicious by the sequencer, it should be quarantined again.

In some implementations of EVM clients, including Geth, transactions can also be removed from the mempool because the accounts that submitted them choose to replace them with different transactions (having the same nonce, but higher gas price). Such replacement transactions will have different hash. Upon replacement, the transaction should continue being quarantined. The sequencer should decide if the replacement transaction should also be quarantined, or if it can be included in the block immediately. The inclusion will result in the quarantine transaction meeting the Nonce release criterion.

\subsection{Transaction Execution}\label{sec:transaction-execution}
Upon releasing from the quarantine, the sequencer can consider including the transaction when forming the next block. 
This is subject to the regular sequencing rules. 
Notably, the transaction may become underpriced in terms of the base fee \cite{eip1559md}. 
In such a case, the sequencer proceeds using its normal sequencing rules—the Ethereum protocol for the mempool management (also used by Optimism Bedrock) keeps the transaction in the mempool until it meets the minimum required base fee. 
If this happens before the transaction retires from the mempool, it can be included in a block. The account that submitted the transaction initially can also desire to resubmit the transaction with a base fee matching the current state of the chain. 
In such a case, the transaction should not be quarantined again. As the modified gas price and limit information would result in a new transaction hash, the SLS protocol has to use the account address, and the transaction data (the function selector and call data), and value, to establish whether a newly incoming transactions is a duplicate of a transaction that has been already released from the quarantine. 

When included in the block, the transaction may succeed or result in a revert and a failure (see the Failure criterion in Section \ref{sec:quarantine}). Both outcomes are acceptable to the SLS protocol.

\section{Implementation}\label{sec:implementation}

In this section, we describe how the SLS protocol can be implemented in the context of the Optimism rollup. 
We first offer a comprehensive overview of the Optimism protocol, followed by a detailed discussion of the modifications required to integrate the SLS protocol. The implementation of SLS is demonstrated on a prototype, which we refer to as \name.


\subsection{Geth and Op-geth}

Geth is an official command-line interface client for Ethereum. 
Written in Go, it serves as a gateway for developers and users to interact with the Ethereum blockchain. 
As a full Ethereum node implementation, Geth plays a pivotal role in the Ethereum ecosystem, allowing for the full realization of its decentralized nature by enabling users to not only participate in the network but also to verify all operations independently.

\texttt{op-geth}, on the other hand, represents a specialized variant of Geth designed to operate within the Optimism Layer 2 scaling solution for Ethereum. Optimism aims to increase Ethereum's transaction throughput and reduce fees while maintaining security and decentralization.
\texttt{op-geth} inherits the core functionalities of Geth, ensuring that developers familiar with Ethereum can easily transition to building and interacting with applications on Optimism.

In the subsequent parts of this section, whenever we introduce a component or feature present in Geth, the same component or feature is also found in \texttt{op-geth}, unless specifically noted otherwise.

\subsection{Queued and pending transactions in Geth}
Geth collects transactions that are used to make new blocks. A block is made at every predefined interval of time called ``block time''. 
During the block time, users submit transactions to Geth that collects them in the transactions pool. The subset of transactions to be included in the next block depends on a set of rules that filter all the transactions in the transactions pool. Specifically, transactions have to go through the queued and pending pools to be accepted in a block. The queued pool is used to keep in memory a list of all the transactions submitted by the users. To get in this queue, transactions have to pass a basic validation that checks the balance, the nonce and some heuristics that are local to the node such as price and size. During this process, if the queue is full, the underpriced transactions are discarded. The transactions that are eventually included in a block are taken from the pending queue. To get in this queue, transactions have to pass a validation that checks if the nonce is too low and if the transaction is underpriced (low balance or too costly). At the end of this process, if the total amount of transactions in the pending queue is over the limit, the highest nonced transactions are removed.

\subsection{Deposit Transactions and Deterministic Chain Derivation}
Optimism does not have a consensus protocol, it instead relies on the L1 to decide on the finality of the L2 blocks. To accomplish this, transactions data is submitted to the L1 in form of L1 transactions' calldata. 
\texttt{op-geth} and retrieved from the L1 to form again the L2 chain in a process called derivation. This mechanism is one of the implementations of a broader concept called "Data Availability"\cite{dataavailability}. Every node of the optimism network can derive the entire chain by just being in connection with an L1 archive node.

The process of derivation consists in collecting all the transactions submitted by the batcher to the batch inbox address and all the user deposited transactions (see Section~\ref{sec:optimism}). 
The collected data is used to reconstruct the L2 blockchain and compute its current state. 
The user deposited transactions are not only used for inter-chain communication, but are also a censorship resistance mechanism introduced by Optimism. 
According to the protocol specification~\cite{optimismdeposits}, deposit transactions cannot be excluded from the blockchain and must be included at the beginning of the first block of every L2 epoch. In practice, if the sequencer excludes these transactions, the result of the computation of the output roots would be successfully challenged in the dispute game.

Deposit transactions can be retrieved from the L1 at any time just like the L2 transactions data submitted by the batcher. Including them in the batches would be of no use and it would consume gas. As a result, in Optimism, deposit transactions are not included in batches. 

In the Optimism protocol, deposit transactions are uncensorable transactions that can perform any action a regular L2 transaction can perform on the L2. This includes malicious actions. For this reason, to allow the SLS to quarantine all the possibly malicious transactions (including deposits), the derivation process needs to be adapted.
Deposit transactions are included into the L2 chain by the \texttt{op-node} service, which reads them from the L1 and in turn provides them to the \texttt{op-geth} which includes them in a block.

\subsection{Detection, Exclusion, and Derivation} 
As described earlier, the malicious transactions detection must happen for both regular and deposit transactions. The regular L2 transactions can be analysed while the selection for inclusion in the pending queue as that's the instant they become valid to be included in a block. For the deposit transactions, they can be analysed during the execution of the \texttt{forkchoiceUpdated} method called by \texttt{op-node}, as that's the instant they are included in a block. The analysis of the transactions can result in a quarantine for some of them. This process is perfectly compatible with the protocol and allows for the building of the block in the case of regular L2 transactions.

For deposit transactions, the quarantining would result in an invalid derivation process. For this reason, we introduced a modification to the protocol that, together with transaction data, submits to the L1 a bitmap where every positive bit represent the inclusion in the block of the deposit transactions. The derivation protocol is updated accordingly allowing other nodes to correctly derive the chain.

\subsection{Deposit Transaction Exclusion}
Deposit transactions are transactions submitted for inclusion into L2 blocks through L1. 
While the Optimism protocol uses such transactions for bridging Ether onto L2, the transactions may have nothing to do with depositing Ether onto the rollup—these can be arbitrary transactions~\cite{optimismdeposits}.
Therefore, they need to be subjected to the same assessment procedure as transactions submitted directly to the rollup mempool. 

Recall that the Optimism state derivation protocol stipulates that deposit transactions have to be included at the beginning of the L2 epoch, i.e., the set of L2 blocks linked to an L1 block in which they were submitted. Quarantining deposit transactions for an arbitrary period of time contradicts this requirement. Furthermore, as the placement of the deposit transactions in the L2 chain is determined, the deposit transactions are not included in batches submitted to L1 by the batcher.
Therefore, in order to allow for a deterministic re-derivation of the state from L1 data only, the SLS protocol needs to publish the following information onto L1 the information about which deposit transactions were accepted by the sequencer, and which deposit transactions were quarantined. 
Furthermore, the protocol would need to publish information about when the deposit transactions were released from the quarantine, and in which L2 blocks they were finally included.

The caveat with quarantined transactions is that after their release from the quarantine, they may fail (see Sections \ref{sec:quarantine} and \ref{sec:transaction-execution}). 
This is not acceptable for the deposit transactions as the deposit transactions may have assets attached to them, and without the ability to complete the deposit on L2, these assets might become permanently locked. 
Therefore, deposit transactions that are deemed malicious by the sequencer enter the quarantine, but are never released. 
The assets attached to these transactions may be released back to the transaction submitter. 

In order to communicate the selection of deposits that should be included in blocks on L2 for the purpose of deterministic chain derivation, the SLS protocol adds a bitmap to the data posted onto the data availability layer. The order of deposit transactions submitted within an L1 block is fixed by the order of transactions in that block. The bitmap contains a 1 bit at the indices matching the deposits to be included, and 0 on indices matching the deposits that are being refused. The number of transactions is capped by the limit on the L1 block size. A single word on Ethereum includes 256 bits, which is enough to cover 256 deposit transactions in a single L1 block. The protocol can use as many words as required to encompass the information about the inclusion of all deposits that fit into a single L1 block. 

The assets locked in smart contracts on L1 due to the refused deposit transactions can be made available for withdrawal back to the accounts that submitted the refused transactions directly on L1 (See details in Section~\ref{subsec:escapehatches}).


\subsection{Risks} \label{subsec:risks}
One major concern for SLS designs is a Denial-of-Service (DoS) attack.
In particular, malicious actors may attempt to flood the system with numerous harmful transactions, leading to entire L2 blocks being void as all transactions processed by the sequencer are quarantined.
However, after SLS quarantines all harmful transactions, the sequencer will continue to process the remaining transactions. As mentioned in Section~\ref{sec:malice-detection}, the sequencer will only perform basic retirement criterion checks on quarantined transactions, e.g. checking transaction expiration, which is very lightweight computation and will not block the sequencer from processing the remaining transactions.

Additionally, performing such an attack would be prohibitively expensive. It requires the malicious party to control a significant number of funded accounts, ensuring that a substantial portion of transactions in the mempool originates from them. If an actor has the resources to execute this strategy with an SLS-enabled sequencer, they could similarly manipulate a standard sequencer by initiating transactions that are certainly to be reverted. Therefore, this type of risk is not unique to SLS-enabled systems.


\section{Discussion}\label{sec:discussion}

In this section, we explore the broader implications of our findings on the SLS protocol, delving into the potential impacts, challenges, and future directions.

\subsection{SLS Implications on Trust Model}

While building the SLS protocol, we made some basic assumptions. One assumption is that the sequencer is centrally operated and trusted. 
In our analysis, we argued that augmenting such a sequencer with SLS does not necessitate an increased level of trust from users towards the sequencer.
This holds even for the deposit transactions. 
Although the vanilla OP stack sequencing protocol stipulates how these transactions need to be processed, it is really up to the honesty of the sequencer if these transactions will or will not be included in the L2 chain. 
An SLS-enabled sequencer explicitly does not process deposit transactions that it deems malicious.

The trust assumptions are not fundamentally increased even by the presence of detectors whose purpose is to identify and segregate malicious transactions—these detectors are operated by the sequencer which is already centralized and assumed to be trusted.

If the SLS-enhanced sequencer implements the Time criterion for quarantine release (see Section \ref{sec:quarantine}), the censorship resistance of the sequencing protocol does not decrease either. 
A centrally operated sequencer can already choose to not process transactions submitted by certain accounts. 
If the Time criterion for quarantine release is implemented, at worst, some transactions may encounter a processing delay. The Time criterion also ensures that transactions that were falsely identified as malicious will eventually get processed too.

\subsection{SLS Implications on Escape Hatches}  \label{subsec:escapehatches}

One of the core risks of a rollup is the ability to withdraw funds from the relevant smart contracts even in the event that the sequencer, state proposer, or other parties responsible for operating a rollup are not online. Such functionality is often called an \emph{escape hatch}~\cite{escapehatches}. This functionality should differ for SLS-enabled sequencers in the following way. Funds present in the last posted L2 state do not need to be handled differently: the state is a result of transactions on L2 that were either quarantined and released, or never quarantined. However, funds that are sent via deposit transactions to the L2 may not have their quarantine status known, as the network may have ceased to operate just after a deposit was recorded on the L1. In such a case, an escape hatch could by default use the Time criterion to allow funds in deposit transactions to withdraw from the L1 contract after a timeout. This would refund the funds and prevent the funds from being considered as part of the L2 state. Deposit transactions which have been quarantined prior to a network failure could also be considered safe to withdraw (on the L1), as they never resulted in malicious behaviour on the L2. However, care should be taken to ensure that these transactions are not considered again in the L2 state if rollup operation resumes.

\subsection{Future Works}

\noindent \textbf{SLS Malicious Transaction Detection.} In this work, we assumed that the sequencer can effectively identify malicious transactions. It should be apparent that the performance of an SLS-enabled sequencing and the overarching blockchain intimately depends on this capability. 
In future work, one can focus on the efficacy of such detection both in terms of the false positive and negative rate, as well as the actual computing performance required. Some related works have been done on Ethereum as introduced in Section~\ref{sec:related}. However, these works are analysis on transactions already included in past blocks; the SLS protocol requires the detection of malicious transactions while they are still in the mempool. Whether parallelization of the detection process can be used to speed up the detection process could be an interesting question.

Other malicious transactions could be detected; we list two directions that are feasible.
First, a detection module could also watch L1 transactions that attempt to deposit stolen funds onto L2, and classify those as malicious on L2, even if they are only depositing funds there.
Second, transactions may be classified as malicious if they result in a smart contract state that breaks a specified invariant.
An SLS-enabled sequencer could receive invariants from applications on the L2 that should be enforced, and the sequencer could simulate transactions to see if these are ever violated.
Violating transactions are then considered malicious.
This requires a suitable invariant language but allows application developers to have their own custom definition of malicious activity.
These approraches are not mutually exclusive with the ideas presented in this work.

\noindent \textbf{SLS on Decentralized Sequencer.} SLS also assumes that the sequencer is centrally operated. There has been a lot of research devoted to decentralized sequencers with a good systematic overview provided by \cite{DBLP:journals/corr/abs-2310-03616}. 
It is unclear how the SLS concept would work with a decentralized sequencer. 
In a decentralized setting, the sequencing nodes would need to form consensus on the quarantine state of transactions. Furthermore, a presence of even a single malicious sequencing node could jeopardize the reliability of the entire decentralized SLS system as such a node could choose to freely include malicious transactions in blocks.

\section{Related Work}\label{sec:related}

In this section, we review the existing literature and prior developments in the field of blockchain security and malice detection, highlighting the advancements and identifying the gaps that our research aims to address.

\subsection{Malicious Transaction Detection}\label{subsec:malicious}

Detecting anomalous transactions on blockchain platforms, particularly Ethereum, has been a focal point of research, aiming to enhance the security and integrity of smart contracts. This subsection delves into significant contributions in the field, categorizing them into two primary approaches: program analysis and machine learning.

\noindent \textbf{Program analysis.} 
Initiatives like \textit{TxSpector} \cite{zhang2020txspector} set the stage for bytecode-level analysis of Ethereum transactions to pinpoint attacks, followed by The Eye of Horus~\cite{ferreira2021eye} and Time-Travel Investigation~\cite{wu2022time}, which further enhanced attack detection capabilities on smart contracts and blockchain transactions. 
More recently, Trace2Inv \cite{chen2024demystifying} dynamically inferred invariants from transaction traces and achieved a significant reduction in false positives. 

\noindent \textbf{Machine learning.} Emerging research by Gai et al.~\cite{gai2023blockchain} and Boxin~\cite{boxin2024research} has explored using machine learning to dynamically detect and mitigate smart contract vulnerabilities, showcasing the potential of AI in securing blockchain ecosystems.

These methodologies, pivotal for identifying malicious activities, complement our Sequencer Level Security (SLS) protocol. By incorporating such techniques, the SLS protocol can effectively quarantine suspicious transactions, leveraging the collective insights from both program analysis and machine learning to ensure robust security within a rollup blockchain.

\subsection{MEV and Frontrunning}
MEV, originally standing for \emph{miner extractable value}, was a concept introduced in the context of proof-of-work (PoW) blockchains. In PoW, miners have the power to decide the order and inclusion of transactions within a block. This setup allows miners to profit by manipulating transaction order~\cite{daian2020flash}. The topic of MEV has attracted significant research interest, with studies aiming to quantify its effects and understand its implications for the blockchain ecosystem~\cite{qin2022quantifying, eskandari2020sok, torres2021frontrunner}.

With Ethereum's transition to proof-of-stake (PoS) in late 2022, the concept of value extraction underwent a redefinition. MEV was reinterpreted as \emph{Maximal Extractable Value}, broadening its scope to encompass any network participant capable of extracting value from transaction manipulation, not just miners.

To address the challenges posed by MEV, the Flashbots organization proposed a novel solution: a bundle mechanism that enables a sealed-bid auction for transaction re-ordering~\cite{obadia2020flashbots, flashbots}. This approach democratizes the MEV extraction process, enabling any actor to compete for the inclusion of transactions in a block.

Although MEV extraction is predominantly profit-driven~\cite{li2023demystifying}, some researchers have explored its potential for thwarting frontrunning attacks. 
Strategies include preempting hackers by inserting transactions that either pause the affected protocol or copy the attacker's transaction to take the funds first before returning them to their rightful owners~\cite{deng2023robust, zhang2023combatting}. A notable instance in July 2023 saw an MEV bot operator successfully intercept and return assets of approximately \$5.5 million USD to a victim of hacking~\cite{frontrungood}.

Different from MEV, our SLS protocol is designed with a singular focus on enhancing security, devoid of profit motives. It does not re-order transactions for profit but rather quarantines suspicious transactions for scrutiny. A transaction identified as malicious will only be processed if it meets specific release criteria, eliminating the need for building external frontrunning bots to monitor and protect target protocols. The SLS protocol provides an integrated solution for monitoring and managing potentially harmful transactions, setting a new standard for security within the blockchain domain.

\section{Conclusion}\label{sec:conclusion}

We described the Sequencer Level Security (SLS) protocol, a novel enhancement protocol for sequencers in rollups. 
This work represents the inaugural exploration of the concept and capability in question, marking it as a pioneering contribution to the field. 
We presented the results specifically for the OP stack on  Ethereum, but our results can be easily generalized and can apply in other settings as well. 
Although we designed the SLS protocol with focus on security, it can be applied for other filtering purposes too. 

The SLS protocol is a novel concept that is currently not implemented in any rollup. We suggest that the existence of SLS-enabled can open a new paradigm for blockchain security where blockchain themselves offer security measures to the resident smart contracts, applications, and users. 
As such, we believe that it will bring opportunities for new research streams in the blockchain security field.

\bibliographystyle{ACM-Reference-Format}
\bibliography{references}


\begin{thebibliography}{52}


\ifx \showCODEN    \undefined \def \showCODEN     #1{\unskip}     \fi
\ifx \showDOI      \undefined \def \showDOI       #1{#1}\fi
\ifx \showISBNx    \undefined \def \showISBNx     #1{\unskip}     \fi
\ifx \showISBNxiii \undefined \def \showISBNxiii  #1{\unskip}     \fi
\ifx \showISSN     \undefined \def \showISSN      #1{\unskip}     \fi
\ifx \showLCCN     \undefined \def \showLCCN      #1{\unskip}     \fi
\ifx \shownote     \undefined \def \shownote      #1{#1}          \fi
\ifx \showarticletitle \undefined \def \showarticletitle #1{#1}   \fi
\ifx \showURL      \undefined \def \showURL       {\relax}        \fi
\providecommand\bibfield[2]{#2}
\providecommand\bibinfo[2]{#2}
\providecommand\natexlab[1]{#1}
\providecommand\showeprint[2][]{arXiv:#2}

\bibitem[get({[n.\,d.]})]%
        {geth}
 \bibinfo{year}{[n.\,d.]}\natexlab{}.
\newblock \bibinfo{title}{{Go-Ethereum}}.
\newblock
\newblock
\urldef\tempurl%
\url{https://geth.ethereum.org/}
\showURL{%
\tempurl}


\bibitem[l2b({[n.\,d.]})]%
        {l2beat}
 \bibinfo{year}{[n.\,d.]}\natexlab{}.
\newblock \bibinfo{title}{{L2BEAT}}.
\newblock
\newblock
\urldef\tempurl%
\url{https://l2beat.com/}
\showURL{%
\tempurl}


\bibitem[dis({[n.\,d.]})]%
        {disputegame}
 \bibinfo{year}{[n.\,d.]}\natexlab{}.
\newblock \bibinfo{title}{Optimism Glossary - {L1} Attributes Deposited Transaction}.
\newblock
\newblock
\urldef\tempurl%
\url{https://github.com/ethereum-optimism/optimism/blob/develop/specs/fault-dispute-game.md}
\showURL{%
\tempurl}


\bibitem[dat({[n.\,d.]})]%
        {dataavailability}
 \bibinfo{year}{[n.\,d.]}\natexlab{}.
\newblock \bibinfo{title}{What is the data availability layer?}
\newblock
\newblock
\urldef\tempurl%
\url{https://www.alchemy.com/overviews/data-availability-layer}
\showURL{%
\tempurl}


\bibitem[bin(2024)]%
        {binancehack}
 \bibinfo{year}{2024}\natexlab{}.
\newblock \bibinfo{title}{Binance Blockchain Back in Action Following \$100M Cross-Chain Bridge Hack}.
\newblock
\newblock
\urldef\tempurl%
\url{https://www.pymnts.com/cryptocurrency/2024/ripple-ceo-crypto-market-will-exceed-5-trillion-by-years-end/}
\showURL{%
\tempurl}


\bibitem[opt(2024a)]%
        {optimismdeposits}
 \bibinfo{year}{2024}\natexlab{a}.
\newblock \bibinfo{title}{Deposited Transactions}.
\newblock
\newblock
\urldef\tempurl%
\url{https://github.com/ethereum-optimism/specs/blob/main/specs/protocol/deposits.md}
\showURL{%
\tempurl}


\bibitem[fla(2024)]%
        {flashbots}
 \bibinfo{year}{2024}\natexlab{}.
\newblock \bibinfo{title}{Flashbots}.
\newblock
\newblock
\urldef\tempurl%
\url{https://www.flashbots.net/}
\showURL{%
\tempurl}


\bibitem[l2b(2024a)]%
        {l2beatArbitrum}
 \bibinfo{year}{2024}\natexlab{a}.
\newblock \bibinfo{title}{{L2BEAT} Arbitrum}.
\newblock
\newblock
\urldef\tempurl%
\url{https://l2beat.com/scaling/projects/arbitrum#risk-analysis}
\showURL{%
\tempurl}


\bibitem[l2b(2024b)]%
        {l2beatBase}
 \bibinfo{year}{2024}\natexlab{b}.
\newblock \bibinfo{title}{{L2BEAT} Base}.
\newblock
\newblock
\urldef\tempurl%
\url{https://l2beat.com/scaling/projects/base#risk-analysis}
\showURL{%
\tempurl}


\bibitem[l2b(2024c)]%
        {l2beatBlast}
 \bibinfo{year}{2024}\natexlab{c}.
\newblock \bibinfo{title}{{L2BEAT} Blast}.
\newblock
\newblock
\urldef\tempurl%
\url{https://l2beat.com/scaling/projects/blast#risk-analysis}
\showURL{%
\tempurl}


\bibitem[l2b(2024d)]%
        {l2beattvl}
 \bibinfo{year}{2024}\natexlab{d}.
\newblock \bibinfo{title}{L2BEAT FAQ}.
\newblock
\newblock
\urldef\tempurl%
\url{https://l2beat.com/faq#why-does-the-total-value-locked-tvl-on-l2beat-differ-from-the-one-on-defillama}
\showURL{%
\tempurl}


\bibitem[l2b(2024e)]%
        {l2beatOptimism}
 \bibinfo{year}{2024}\natexlab{e}.
\newblock \bibinfo{title}{{L2BEAT} Optimism}.
\newblock
\newblock
\urldef\tempurl%
\url{https://l2beat.com/scaling/projects/optimism#risk-analysis}
\showURL{%
\tempurl}


\bibitem[l2b(2024f)]%
        {l2beatStarknet}
 \bibinfo{year}{2024}\natexlab{f}.
\newblock \bibinfo{title}{{L2BEAT} Starknet}.
\newblock
\newblock
\urldef\tempurl%
\url{https://l2beat.com/scaling/projects/starknet#risk-analysis}
\showURL{%
\tempurl}


\bibitem[l2b(2024g)]%
        {l2beatZkSyncEra}
 \bibinfo{year}{2024}\natexlab{g}.
\newblock \bibinfo{title}{{L2BEAT} zkSync Era}.
\newblock
\newblock
\urldef\tempurl%
\url{https://l2beat.com/scaling/projects/zksync-era#risk-analysis}
\showURL{%
\tempurl}


\bibitem[mpt(2024)]%
        {mpt}
 \bibinfo{year}{2024}\natexlab{}.
\newblock \bibinfo{title}{Merkle Patricia Trie}.
\newblock
\newblock
\urldef\tempurl%
\url{https://ethereum.org/en/developers/docs/data-structures-and-encoding/patricia-merkle-trie/}
\showURL{%
\tempurl}


\bibitem[fro(2024)]%
        {frontrungood}
 \bibinfo{year}{2024}\natexlab{}.
\newblock \bibinfo{title}{MEV bot runner `c0ffeebabe.eth' returns \$5.4 million amid Curve exploit}.
\newblock
\newblock
\urldef\tempurl%
\url{https://www.flashbots.net/}
\showURL{%
\tempurl}


\bibitem[cen(2024)]%
        {centralizedrollup}
 \bibinfo{year}{2024}\natexlab{}.
\newblock \bibinfo{title}{Rollup Sequencers Are Centralized — And That's Fine}.
\newblock
\newblock
\urldef\tempurl%
\url{https://blockworks.co/news/rollup-sequencers-are-centralized}
\showURL{%
\tempurl}


\bibitem[opt(2024b)]%
        {optimismchallenge}
 \bibinfo{year}{2024}\natexlab{b}.
\newblock \bibinfo{title}{Understanding the Challenge Period}.
\newblock
\newblock
\urldef\tempurl%
\url{https://docs.optimism.io/builders/app-developers/bridging/messaging}
\showURL{%
\tempurl}


\bibitem[bit(2024)]%
        {bitcoinincident}
 \bibinfo{year}{2024}\natexlab{}.
\newblock \bibinfo{title}{Value overflow incident}.
\newblock
\newblock
\urldef\tempurl%
\url{https://en.bitcoin.it/wiki/Value_overflow_incident}
\showURL{%
\tempurl}


\bibitem[eth(2024)]%
        {ethereumdao}
 \bibinfo{year}{2024}\natexlab{}.
\newblock \bibinfo{title}{What Was The DAO?}
\newblock
\newblock
\urldef\tempurl%
\url{https://www.gemini.com/cryptopedia/the-dao-hack-makerdao#section-origins-of-the-dao}
\showURL{%
\tempurl}


\bibitem[Back et~al\mbox{.}(2014)]%
        {back2014enabling}
\bibfield{author}{\bibinfo{person}{Adam Back}, \bibinfo{person}{Matt Corallo}, \bibinfo{person}{Luke Dashjr}, \bibinfo{person}{Mark Friedenbach}, \bibinfo{person}{Gregory Maxwell}, \bibinfo{person}{Andrew Miller}, \bibinfo{person}{Andrew Poelstra}, \bibinfo{person}{Jorge Tim{\'o}n}, {and} \bibinfo{person}{Pieter Wuille}.} \bibinfo{year}{2014}\natexlab{}.
\newblock \showarticletitle{Enabling blockchain innovations with pegged sidechains}.
\newblock   \bibinfo{volume}{72} (\bibinfo{year}{2014}), \bibinfo{pages}{201--224}.
\newblock
\urldef\tempurl%
\url{http://www.opensciencereview.com/papers/123/enablingblockchain-innovations-with-pegged-sidechains}
\showURL{%
\tempurl}


\bibitem[Boxin(2024)]%
        {boxin2024research}
\bibfield{author}{\bibinfo{person}{Yang Boxin}.} \bibinfo{year}{2024}\natexlab{}.
\newblock \showarticletitle{Research on Dynamic Detection of Vulnerabilities in Smart Contracts Based on Machine Learning}. In \bibinfo{booktitle}{\emph{2024 IEEE 3rd International Conference on Electrical Engineering, Big Data and Algorithms (EEBDA)}}. IEEE, \bibinfo{pages}{219--223}.
\newblock
\newblock
\shownote{To appear.}.


\bibitem[Buterin(2022a)]%
        {eip1559md}
\bibfield{author}{\bibinfo{person}{Vitalik Buterin}.} \bibinfo{year}{2022}\natexlab{a}.
\newblock \bibinfo{title}{Multidimensional {EIP} 1559}.
\newblock
\newblock
\urldef\tempurl%
\url{https://ethresear.ch/t/multidimensional-eip-1559/11651}
\showURL{%
\tempurl}


\bibitem[Buterin(2022b)]%
        {eip4844-protodanksharding}
\bibfield{author}{\bibinfo{person}{Vitalik Buterin}.} \bibinfo{year}{2022}\natexlab{b}.
\newblock \bibinfo{title}{Proto-Danksharding FAQ}.
\newblock
\newblock
\urldef\tempurl%
\url{https://notes.ethereum.org/@vbuterin/proto_danksharding_faq}
\showURL{%
\tempurl}


\bibitem[Chen et~al\mbox{.}(2024)]%
        {chen2024demystifying}
\bibfield{author}{\bibinfo{person}{Zhiyang Chen}, \bibinfo{person}{Ye Liu}, \bibinfo{person}{Sidi~Mohamed Beillahi}, \bibinfo{person}{Yi Li}, {and} \bibinfo{person}{Fan Long}.} \bibinfo{year}{2024}\natexlab{}.
\newblock \showarticletitle{Demystifying Invariant Effectiveness for Securing Smart Contracts}.
\newblock  (\bibinfo{year}{2024}).
\newblock
\newblock
\shownote{arXiv preprint arXiv:2404.14580. Forthcoming in the Proceedings of the 2024 31th Joint Meeting on Foundations of Software Engineering (FSE 2024)}.


\bibitem[Daian et~al\mbox{.}(2020)]%
        {daian2020flash}
\bibfield{author}{\bibinfo{person}{Philip Daian}, \bibinfo{person}{Steven Goldfeder}, \bibinfo{person}{Tyler Kell}, \bibinfo{person}{Yunqi Li}, \bibinfo{person}{Xueyuan Zhao}, \bibinfo{person}{Iddo Bentov}, \bibinfo{person}{Lorenz Breidenbach}, {and} \bibinfo{person}{Ari Juels}.} \bibinfo{year}{2020}\natexlab{}.
\newblock \showarticletitle{Flash Boys 2.0: Frontrunning in Decentralized Exchanges, Miner Extractable Value, and Consensus Instability}. In \bibinfo{booktitle}{\emph{2020 {IEEE} Symposium on Security and Privacy, {SP} 2020, San Francisco, CA, USA, May 18-21, 2020}}. \bibinfo{publisher}{{IEEE}}, \bibinfo{pages}{910--927}.
\newblock
\urldef\tempurl%
\url{https://doi.org/10.1109/SP40000.2020.00040}
\showDOI{\tempurl}


\bibitem[Deng et~al\mbox{.}(2023)]%
        {deng2023robust}
\bibfield{author}{\bibinfo{person}{Xun Deng}, \bibinfo{person}{Zihan Zhao}, \bibinfo{person}{Sidi~Mohamed Beillahi}, \bibinfo{person}{Han Du}, \bibinfo{person}{Cyrus Minwalla}, \bibinfo{person}{Keerthi Nelaturu}, \bibinfo{person}{Andreas~G. Veneris}, {and} \bibinfo{person}{Fan Long}.} \bibinfo{year}{2023}\natexlab{}.
\newblock \showarticletitle{A Robust Front-Running Methodology for Malicious Flash-Loan {DeFi} Attacks}. In \bibinfo{booktitle}{\emph{{IEEE} International Conference on Decentralized Applications and Infrastructures, {DAPPS} 2023, Athens, Greece, July 17-20, 2023}}. \bibinfo{publisher}{{IEEE}}, \bibinfo{pages}{38--47}.
\newblock
\urldef\tempurl%
\url{https://doi.org/10.1109/DAPPS57946.2023.00015}
\showDOI{\tempurl}


\bibitem[Eskandari et~al\mbox{.}(2019)]%
        {eskandari2020sok}
\bibfield{author}{\bibinfo{person}{Shayan Eskandari}, \bibinfo{person}{Seyedehmahsa Moosavi}, {and} \bibinfo{person}{Jeremy Clark}.} \bibinfo{year}{2019}\natexlab{}.
\newblock \showarticletitle{{SoK}: Transparent Dishonesty: Front-Running Attacks on Blockchain}. In \bibinfo{booktitle}{\emph{Financial Cryptography and Data Security - {FC} 2019 International Workshops, {VOTING} and WTSC, St. Kitts, St. Kitts and Nevis, February 18-22, 2019, Revised Selected Papers}} \emph{(\bibinfo{series}{Lecture Notes in Computer Science}, Vol.~\bibinfo{volume}{11599})}, \bibfield{editor}{\bibinfo{person}{Andrea Bracciali}, \bibinfo{person}{Jeremy Clark}, \bibinfo{person}{Federico Pintore}, \bibinfo{person}{Peter~B. R{\o}nne}, {and} \bibinfo{person}{Massimiliano Sala}} (Eds.). \bibinfo{publisher}{Springer}, \bibinfo{pages}{170--189}.
\newblock
\urldef\tempurl%
\url{https://doi.org/10.1007/978-3-030-43725-1\_13}
\showDOI{\tempurl}


\bibitem[Gai et~al\mbox{.}(2023)]%
        {gai2023blockchain}
\bibfield{author}{\bibinfo{person}{Yu Gai}, \bibinfo{person}{Liyi Zhou}, \bibinfo{person}{Kaihua Qin}, \bibinfo{person}{Dawn Song}, {and} \bibinfo{person}{Arthur Gervais}.} \bibinfo{year}{2023}\natexlab{}.
\newblock \showarticletitle{Blockchain Large Language Models}.
\newblock \bibinfo{journal}{\emph{CoRR}}  \bibinfo{volume}{abs/2304.12749} (\bibinfo{year}{2023}).
\newblock
\urldef\tempurl%
\url{https://doi.org/10.48550/ARXIV.2304.12749}
\showDOI{\tempurl}
\showeprint[arXiv]{2304.12749}


\bibitem[Gorzny et~al\mbox{.}(2022)]%
        {escapehatches}
\bibfield{author}{\bibinfo{person}{Jan Gorzny}, \bibinfo{person}{Lin Po-An}, {and} \bibinfo{person}{Martin Derka}.} \bibinfo{year}{2022}\natexlab{}.
\newblock \showarticletitle{Ideal Properties of Rollup Escape Hatches}. In \bibinfo{booktitle}{\emph{Proceedings of the 3rd International Workshop on Distributed Infrastructure for the Common Good}} (Quebec, Quebec City, Canada) \emph{(\bibinfo{series}{DICG '22})}. \bibinfo{publisher}{Association for Computing Machinery}, \bibinfo{address}{New York, NY, USA}, \bibinfo{pages}{7–12}.
\newblock
\showISBNx{9781450399289}
\urldef\tempurl%
\url{https://doi.org/10.1145/3565383.3566107}
\showDOI{\tempurl}


\bibitem[Gudgeon et~al\mbox{.}(2020)]%
        {GudgeonMRMG20}
\bibfield{author}{\bibinfo{person}{Lewis Gudgeon}, \bibinfo{person}{Pedro Moreno{-}Sanchez}, \bibinfo{person}{Stefanie Roos}, \bibinfo{person}{Patrick McCorry}, {and} \bibinfo{person}{Arthur Gervais}.} \bibinfo{year}{2020}\natexlab{}.
\newblock \showarticletitle{{SoK}: Layer-Two Blockchain Protocols}. In \bibinfo{booktitle}{\emph{Financial Cryptography and Data Security - 24th International Conference, {FC} 2020, Kota Kinabalu, Malaysia, February 10-14, 2020 Revised Selected Papers}} \emph{(\bibinfo{series}{Lecture Notes in Computer Science}, Vol.~\bibinfo{volume}{12059})}, \bibfield{editor}{\bibinfo{person}{Joseph Bonneau} {and} \bibinfo{person}{Nadia Heninger}} (Eds.). \bibinfo{publisher}{Springer}, \bibinfo{pages}{201--226}.
\newblock
\urldef\tempurl%
\url{https://doi.org/10.1007/978-3-030-51280-4\_12}
\showDOI{\tempurl}


\bibitem[Kane and Heinrich(1992)]%
        {kane1992mips}
\bibfield{author}{\bibinfo{person}{Gerry Kane} {and} \bibinfo{person}{Joe Heinrich}.} \bibinfo{year}{1992}\natexlab{}.
\newblock \bibinfo{booktitle}{\emph{MIPS RISC architectures}}.
\newblock \bibinfo{publisher}{Prentice-Hall, Inc.}
\newblock


\bibitem[Khalil et~al\mbox{.}(2018)]%
        {cryptoeprint:2018/642}
\bibfield{author}{\bibinfo{person}{Rami Khalil}, \bibinfo{person}{Alexei Zamyatin}, \bibinfo{person}{Guillaume Felley}, \bibinfo{person}{Pedro Moreno-Sanchez}, {and} \bibinfo{person}{Arthur Gervais}.} \bibinfo{year}{2018}\natexlab{}.
\newblock \bibinfo{title}{Commit-Chains: Secure, Scalable Off-Chain Payments}.
\newblock \bibinfo{howpublished}{Cryptology ePrint Archive, Paper 2018/642}.
\newblock
\urldef\tempurl%
\url{https://eprint.iacr.org/2018/642}
\showURL{%
\tempurl}


\bibitem[Li et~al\mbox{.}(2023)]%
        {li2023demystifying}
\bibfield{author}{\bibinfo{person}{Zihao Li}, \bibinfo{person}{Jianfeng Li}, \bibinfo{person}{Zheyuan He}, \bibinfo{person}{Xiapu Luo}, \bibinfo{person}{Ting Wang}, \bibinfo{person}{Xiaoze Ni}, \bibinfo{person}{Wenwu Yang}, \bibinfo{person}{Xi Chen}, {and} \bibinfo{person}{Ting Chen}.} \bibinfo{year}{2023}\natexlab{}.
\newblock \showarticletitle{Demystifying {DeFi} {MEV} Activities in Flashbots Bundle}. In \bibinfo{booktitle}{\emph{Proceedings of the 2023 {ACM} {SIGSAC} Conference on Computer and Communications Security, {CCS} 2023, Copenhagen, Denmark, November 26-30, 2023}}, \bibfield{editor}{\bibinfo{person}{Weizhi Meng}, \bibinfo{person}{Christian~Damsgaard Jensen}, \bibinfo{person}{Cas Cremers}, {and} \bibinfo{person}{Engin Kirda}} (Eds.). \bibinfo{publisher}{{ACM}}, \bibinfo{pages}{165--179}.
\newblock
\urldef\tempurl%
\url{https://doi.org/10.1145/3576915.3616590}
\showDOI{\tempurl}


\bibitem[McCorry et~al\mbox{.}(2021)]%
        {cryptoeprint:2021/1589}
\bibfield{author}{\bibinfo{person}{Patrick McCorry}, \bibinfo{person}{Chris Buckland}, \bibinfo{person}{Bennet Yee}, {and} \bibinfo{person}{Dawn Song}.} \bibinfo{year}{2021}\natexlab{}.
\newblock \bibinfo{title}{{SoK}: Validating Bridges as a Scaling Solution for Blockchains}.
\newblock \bibinfo{howpublished}{Cryptology ePrint Archive, Paper 2021/1589}.
\newblock
\urldef\tempurl%
\url{https://eprint.iacr.org/2021/1589}
\showURL{%
\tempurl}
\newblock
\shownote{\url{https://eprint.iacr.org/2021/1589}}.


\bibitem[Meyerson(2014)]%
        {meyerson2014go}
\bibfield{author}{\bibinfo{person}{Jeff Meyerson}.} \bibinfo{year}{2014}\natexlab{}.
\newblock \showarticletitle{The Go Programming Language}.
\newblock \bibinfo{journal}{\emph{{IEEE} Softw.}} \bibinfo{volume}{31}, \bibinfo{number}{5} (\bibinfo{year}{2014}), \bibinfo{pages}{104}.
\newblock
\urldef\tempurl%
\url{https://doi.org/10.1109/MS.2014.127}
\showDOI{\tempurl}


\bibitem[Motepalli et~al\mbox{.}(2023)]%
        {DBLP:journals/corr/abs-2310-03616}
\bibfield{author}{\bibinfo{person}{Shashank Motepalli}, \bibinfo{person}{Luciano Freitas}, {and} \bibinfo{person}{Benjamin Livshits}.} \bibinfo{year}{2023}\natexlab{}.
\newblock \showarticletitle{{SoK}: Decentralized Sequencers for Rollups}.
\newblock \bibinfo{journal}{\emph{CoRR}}  \bibinfo{volume}{abs/2310.03616} (\bibinfo{year}{2023}).
\newblock
\urldef\tempurl%
\url{https://doi.org/10.48550/ARXIV.2310.03616}
\showDOI{\tempurl}
\showeprint[arXiv]{2310.03616}


\bibitem[Nakamoto(2008)]%
        {bitcoin}
\bibfield{author}{\bibinfo{person}{Satoshi Nakamoto}.} \bibinfo{year}{2008}\natexlab{}.
\newblock \bibinfo{title}{Bitcoin: A peer-to-peer electronic cash system}.
\newblock
\newblock
\urldef\tempurl%
\url{https://bitcoin.org/bitcoin.pdf}
\showURL{%
\tempurl}
\newblock
\shownote{\url{https://bitcoin.org/bitcoin.pdf}}.


\bibitem[Nguyen et~al\mbox{.}(2019)]%
        {8746079}
\bibfield{author}{\bibinfo{person}{Cong~Thanh Nguyen}, \bibinfo{person}{Dinh~Thai Hoang}, \bibinfo{person}{Diep~N. Nguyen}, \bibinfo{person}{Dusit Niyato}, \bibinfo{person}{Huynh~Tuong Nguyen}, {and} \bibinfo{person}{Eryk Dutkiewicz}.} \bibinfo{year}{2019}\natexlab{}.
\newblock \showarticletitle{Proof-of-Stake Consensus Mechanisms for Future Blockchain Networks: Fundamentals, Applications and Opportunities}.
\newblock \bibinfo{journal}{\emph{{IEEE} Access}}  \bibinfo{volume}{7} (\bibinfo{year}{2019}), \bibinfo{pages}{85727--85745}.
\newblock
\urldef\tempurl%
\url{https://doi.org/10.1109/ACCESS.2019.2925010}
\showDOI{\tempurl}


\bibitem[Obadia(2020)]%
        {obadia2020flashbots}
\bibfield{author}{\bibinfo{person}{Alex Obadia}.} \bibinfo{year}{2020}\natexlab{}.
\newblock \showarticletitle{Flashbots: Frontrunning the MEV crisis}.
\newblock  (\bibinfo{year}{2020}).
\newblock


\bibitem[Optimism({[n.\,d.]})]%
        {opstack}
\bibfield{author}{\bibinfo{person}{Optimism}.} \bibinfo{year}{[n.\,d.]}\natexlab{}.
\newblock \bibinfo{title}{Cross-rollup {NFT} wrapper and migration ideas}.
\newblock
\newblock
\urldef\tempurl%
\url{https://docs.optimism.io/stack/getting-started}
\showURL{%
\tempurl}
\newblock
\shownote{\url{https://docs.optimism.io/stack/getting-started}}.


\bibitem[Qin et~al\mbox{.}(2022)]%
        {qin2022quantifying}
\bibfield{author}{\bibinfo{person}{Kaihua Qin}, \bibinfo{person}{Liyi Zhou}, {and} \bibinfo{person}{Arthur Gervais}.} \bibinfo{year}{2022}\natexlab{}.
\newblock \showarticletitle{Quantifying Blockchain Extractable Value: How dark is the forest?}. In \bibinfo{booktitle}{\emph{43rd {IEEE} Symposium on Security and Privacy, {SP} 2022, San Francisco, CA, USA, May 22-26, 2022}}. \bibinfo{publisher}{{IEEE}}, \bibinfo{pages}{198--214}.
\newblock
\urldef\tempurl%
\url{https://doi.org/10.1109/SP46214.2022.9833734}
\showDOI{\tempurl}


\bibitem[Schollmeier(2001)]%
        {schollmeier2001definition}
\bibfield{author}{\bibinfo{person}{R{\"{u}}diger Schollmeier}.} \bibinfo{year}{2001}\natexlab{}.
\newblock \showarticletitle{A Definition of Peer-to-Peer Networking for the Classification of Peer-to-Peer Architectures and Applications}. In \bibinfo{booktitle}{\emph{1st International Conference on Peer-to-Peer Computing {(P2P} 2001), 27-29 August 2001, Link{\"{o}}ping, Sweden}}, \bibfield{editor}{\bibinfo{person}{Ross~Lee Graham} {and} \bibinfo{person}{Nahid Shahmehri}} (Eds.). \bibinfo{publisher}{{IEEE} Computer Society}, \bibinfo{pages}{101--102}.
\newblock
\urldef\tempurl%
\url{https://doi.org/10.1109/P2P.2001.990434}
\showDOI{\tempurl}


\bibitem[Szabo(1996)]%
        {szabo1997idea}
\bibfield{author}{\bibinfo{person}{Nick Szabo}.} \bibinfo{year}{1996}\natexlab{}.
\newblock \showarticletitle{Smart Contracts: Building Blocks for Digital Markets}.
\newblock  (\bibinfo{year}{1996}).
\newblock


\bibitem[Torres et~al\mbox{.}(2021a)]%
        {torres2021frontrunner}
\bibfield{author}{\bibinfo{person}{Christof~Ferreira Torres}, \bibinfo{person}{Ramiro Camino}, {and} \bibinfo{person}{Radu State}.} \bibinfo{year}{2021}\natexlab{a}.
\newblock \showarticletitle{Frontrunner {J}ones and the {R}aiders of the {D}ark {F}orest: An Empirical Study of Frontrunning on the {E}thereum Blockchain}. In \bibinfo{booktitle}{\emph{30th {USENIX} Security Symposium, {USENIX} Security 2021, August 11-13, 2021}}, \bibfield{editor}{\bibinfo{person}{Michael~D. Bailey} {and} \bibinfo{person}{Rachel Greenstadt}} (Eds.). \bibinfo{publisher}{{USENIX} Association}, \bibinfo{pages}{1343--1359}.
\newblock
\urldef\tempurl%
\url{https://www.usenix.org/conference/usenixsecurity21/presentation/torres}
\showURL{%
\tempurl}


\bibitem[Torres et~al\mbox{.}(2021b)]%
        {ferreira2021eye}
\bibfield{author}{\bibinfo{person}{Christof~Ferreira Torres}, \bibinfo{person}{Antonio~Ken Iannillo}, \bibinfo{person}{Arthur Gervais}, {and} \bibinfo{person}{Radu State}.} \bibinfo{year}{2021}\natexlab{b}.
\newblock \showarticletitle{The Eye of {H}orus: Spotting and Analyzing Attacks on {E}thereum Smart Contracts}. In \bibinfo{booktitle}{\emph{Financial Cryptography and Data Security - 25th International Conference, {FC} 2021, Virtual Event, March 1-5, 2021, Revised Selected Papers, Part {I}}} \emph{(\bibinfo{series}{Lecture Notes in Computer Science}, Vol.~\bibinfo{volume}{12674})}, \bibfield{editor}{\bibinfo{person}{Nikita Borisov} {and} \bibinfo{person}{Claudia D{\'{\i}}az}} (Eds.). \bibinfo{publisher}{Springer}, \bibinfo{pages}{33--52}.
\newblock
\urldef\tempurl%
\url{https://doi.org/10.1007/978-3-662-64322-8\_2}
\showDOI{\tempurl}


\bibitem[Werner et~al\mbox{.}(2022)]%
        {DBLP:conf/aft/Werner0GKHK22}
\bibfield{author}{\bibinfo{person}{Sam Werner}, \bibinfo{person}{Daniel Perez}, \bibinfo{person}{Lewis Gudgeon}, \bibinfo{person}{Ariah Klages{-}Mundt}, \bibinfo{person}{Dominik Harz}, {and} \bibinfo{person}{William~J. Knottenbelt}.} \bibinfo{year}{2022}\natexlab{}.
\newblock \showarticletitle{{SoK}: Decentralized Finance ({DeFi})}. In \bibinfo{booktitle}{\emph{Proceedings of the 4th {ACM} Conference on Advances in Financial Technologies, {AFT} 2022, Cambridge, MA, USA, September 19-21, 2022}}, \bibfield{editor}{\bibinfo{person}{Maurice Herlihy} {and} \bibinfo{person}{Neha Narula}} (Eds.). \bibinfo{publisher}{{ACM}}, \bibinfo{pages}{30--46}.
\newblock
\urldef\tempurl%
\url{https://doi.org/10.1145/3558535.3559780}
\showDOI{\tempurl}


\bibitem[Wood(2014)]%
        {ethereum}
\bibfield{author}{\bibinfo{person}{Gavin Wood}.} \bibinfo{year}{2014}\natexlab{}.
\newblock \bibinfo{title}{Ethereum: A secure decentralised generalised transaction ledger}.
\newblock
\newblock
\urldef\tempurl%
\url{https://ethereum.github.io/yellowpaper/paper.pdf}
\showURL{%
\tempurl}
\newblock
\shownote{Ethereum Project Yellow Paper}.


\bibitem[Wu et~al\mbox{.}(2022)]%
        {wu2022time}
\bibfield{author}{\bibinfo{person}{Siwei Wu}, \bibinfo{person}{Lei Wu}, \bibinfo{person}{Yajin Zhou}, \bibinfo{person}{Runhuai Li}, \bibinfo{person}{Zhi Wang}, \bibinfo{person}{Xiapu Luo}, \bibinfo{person}{Cong Wang}, {and} \bibinfo{person}{Kui Ren}.} \bibinfo{year}{2022}\natexlab{}.
\newblock \showarticletitle{Time-travel Investigation: Toward Building a Scalable Attack Detection Framework on {E}thereum}.
\newblock \bibinfo{journal}{\emph{{ACM} Trans. Softw. Eng. Methodol.}} \bibinfo{volume}{31}, \bibinfo{number}{3} (\bibinfo{year}{2022}), \bibinfo{pages}{54:1--54:33}.
\newblock
\urldef\tempurl%
\url{https://doi.org/10.1145/3505263}
\showDOI{\tempurl}


\bibitem[Zamani et~al\mbox{.}(2018)]%
        {zamani2018rapidchain}
\bibfield{author}{\bibinfo{person}{Mahdi Zamani}, \bibinfo{person}{Mahnush Movahedi}, {and} \bibinfo{person}{Mariana Raykova}.} \bibinfo{year}{2018}\natexlab{}.
\newblock \showarticletitle{Rapid{C}hain: Scaling Blockchain via Full Sharding}. In \bibinfo{booktitle}{\emph{Proceedings of the 2018 {ACM} {SIGSAC} Conference on Computer and Communications Security, {CCS} 2018, Toronto, ON, Canada, October 15-19, 2018}}, \bibfield{editor}{\bibinfo{person}{David Lie}, \bibinfo{person}{Mohammad Mannan}, \bibinfo{person}{Michael Backes}, {and} \bibinfo{person}{XiaoFeng Wang}} (Eds.). \bibinfo{publisher}{{ACM}}, \bibinfo{pages}{931--948}.
\newblock
\urldef\tempurl%
\url{https://doi.org/10.1145/3243734.3243853}
\showDOI{\tempurl}


\bibitem[Zhang et~al\mbox{.}(2020)]%
        {zhang2020txspector}
\bibfield{author}{\bibinfo{person}{Mengya Zhang}, \bibinfo{person}{Xiaokuan Zhang}, \bibinfo{person}{Yinqian Zhang}, {and} \bibinfo{person}{Zhiqiang Lin}.} \bibinfo{year}{2020}\natexlab{}.
\newblock \showarticletitle{{TXSPECTOR:} Uncovering Attacks in {E}thereum from Transactions}. In \bibinfo{booktitle}{\emph{29th {USENIX} Security Symposium, {USENIX} Security 2020, August 12-14, 2020}}, \bibfield{editor}{\bibinfo{person}{Srdjan Capkun} {and} \bibinfo{person}{Franziska Roesner}} (Eds.). \bibinfo{publisher}{{USENIX} Association}, \bibinfo{pages}{2775--2792}.
\newblock
\urldef\tempurl%
\url{https://www.usenix.org/conference/usenixsecurity20/presentation/zhang-mengya}
\showURL{%
\tempurl}


\bibitem[Zhang et~al\mbox{.}(2023)]%
        {zhang2023combatting}
\bibfield{author}{\bibinfo{person}{Wuqi Zhang}, \bibinfo{person}{Lili Wei}, \bibinfo{person}{Shing{-}Chi Cheung}, \bibinfo{person}{Yepang Liu}, \bibinfo{person}{Shuqing Li}, \bibinfo{person}{Lu Liu}, {and} \bibinfo{person}{Michael~R. Lyu}.} \bibinfo{year}{2023}\natexlab{}.
\newblock \showarticletitle{Combatting Front-Running in Smart Contracts: Attack Mining, Benchmark Construction and Vulnerability Detector Evaluation}.
\newblock \bibinfo{journal}{\emph{{IEEE} Trans. Software Eng.}} \bibinfo{volume}{49}, \bibinfo{number}{6} (\bibinfo{year}{2023}), \bibinfo{pages}{3630--3646}.
\newblock
\urldef\tempurl%
\url{https://doi.org/10.1109/TSE.2023.3270117}
\showDOI{\tempurl}


\end{thebibliography}


\end{document}